\documentstyle[aps,multicol,psfig,epsfig]{revtex}   

\newcommand{\beq}{\begin{equation}}  
\newcommand{\eeq}{\end{equation}}  
\newcommand{\bea}{\begin{eqnarray}}  
\newcommand{\eea}{\end{eqnarray}}  
  
\begin{document}     
\draft     
\title{Pathways to double ionization of atoms in strong fields}
\author{
Krzysztof Sacha$^{1,2}$ and Bruno Eckhardt$^1$
}     
\address{$^1$ Fachbereich Physik, Philipps Universit\"at    
	 Marburg, D-35032 Marburg, Germany}    
\address{$^2$ Instytut Fizyki im. Mariana Smoluchowskiego,
Uniwersytet Jagiello\'nski, ul. Reymonta 4,
PL-30-059 Krak\'ow, Poland}
\date{\today}     
\maketitle{ }   
  
\begin{abstract}
We discuss the final stages of double ionization of atoms in 
a strong linearly polarized laser field within a classical model. 
We propose that all trajectories leading to non-sequential double ionization
pass close to a saddle in phase space which we identify and characterize.
The saddle lies in a two degree of freedom subspace of symmetrically escaping
electrons. The distribution of longitudinal momenta of ions as calculated
within the subspace shows the double hump structure observed in
experiments. Including a symmetric bending mode of the electrons allows us to
reproduce the transverse ion momenta. We discuss also a path to
sequential ionization and show that it does not lead to the observed 
momentum distributions.
\end{abstract}     
  
\begin{multicols}{2}

\section{Introduction}
Present day lasers are powerful enough
to ionize several electrons from an atom. The electrons can be removed 
one by one in a sequential process or all at once in a 
non-sequential process.
Independent electron models give ionization rates
that are much smaller than experimentally observed \cite{lhuiller}, 
indicating that
interactions between electrons are important. 
A series of most recent experiments has added the observation that 
also the final state of the electrons is dominated by the interactions: the
total momentum of the electrons is aligned along the field
axis \cite{weber1,weber2,rottke,weber3} 
and the joint distribution of parallel momenta for the two
electrons, in the double ionization experiment \cite{weber3}, 
has pronounced maxima along the diagonal, showing that 
the electrons typically come with the same momenta. 
These observations have been reproduced in numerical simulations
with varying approximations and simplifying assumptions 
\cite{becker4,wbecker}. But given the 
complexity of the analysis that is required the essential elements are 
difficult to identify. 
As a step towards a better understanding we discuss here
the pathways to double ionization within a classical model for 
electrons in a combined Coulomb and external field. 

Our aim is not to describe the full ionization process all the way
from the ground state to the final, multiply ionized state.
According to the currently accepted models 
\cite{Corkum,Kulander3,becker1,becker2,Kulander5,becker3,becker4,wbecker}, 
the whole process
of multiphoton multiple ionization
can naturally be divided into two steps: in the first step a 
compound state of highly excited electrons close to the nucleus
is formed and in the second step several electrons can escape from 
this compound state to produce the multiply ionized final state.
We focus on the last step, the escape of two or more electrons from
the highly excited compound state close to the nucleus. 

The formation 
of the intermediate compound state is suggested by the rescattering model 
\cite{Corkum,Kulander3} for
strong field multiple ionization. According to this model the enhanced
cross section for multiple ionization is due to a rescattering
of one electron that is temporarily ionized and accelerated by the 
field before it returns to the nucleus when the field reverses. 
During the collision energy is transfered to other electrons,
but all of this happens close to the core, where the dynamics of the 
electrons is fast and the interactions are strong and non-integrable.
As a consequence, details of the initial preparation process are lost.
Moreover, the decay of this state is also quick. 
The compound state is thus a short lived,
highly unstable complex that separates the first half of the excitation
process, whose main contribution is the build up of energy in the complex,
from the second half, where the decay mode is determined.

The compound state has several decay paths: single ionization when only
one electron escapes to infinity, double or multiple ionization with
two or more electrons escaping to infinity, and the case of a single 
electron that escapes from the neighborhood of the nucleus but is rescattered
to the next field reversal: in the latter case the whole process
repeats itself, another compound state is formed and 
the decay path has to be selected anew.

We discuss here the further evolution after the
formation of the compound state. To be specific, we will focus on the 
escape of two electrons in the following, but the arguments can
easily be extended to the removal of more than two electrons. The main
aim is to identify the channels that lead to double escape and to 
study their signatures in the distribution of electron and ion momenta.
Our analysis is purely classical.
Given the highly excited complex from which we start and the multiphoton nature
of the process this seems a reasonable point of entry to the final stage of the
ionization process. 
Our analysis is very similar to Wanniers approach
\cite{Wannier,Rau,Rost,Rost1} to 
double ionization by electron impact.
There main difference is that in
the present case one has to take into account also the external field. 
Brief summaries of some aspects of this model have been presented in
\cite{ES1,ES2}. 

The pathways to double ionization are discussed in section 2. The
dynamics in the $C_{2v}$ and $C_{v}$ subspaces including 
the effective potential
and sample trajectories is discussed in section 3. A key
element of our argument is the identification of a saddle in 
phase space near which trajectories leading to 
non-sequential double ionization have
to pass; its properties and stability are discussed in section 4.
The distributions of electron momenta within the $C_{2v}$ and $C_{v}$
subspaces are analyzed in section 5. The dynamics outside these
symmetry spaces and sequential ionization processes 
are discussed in section 6. We conclude with a
summary of the model in section 7.

\section{Pathways to double ionization}
As described in the introduction we assume an initial state of 
two highly excited electrons near the nucleus which then decays
to either single ionization or double ionization. During this
process the linearly polarized laser field is always on. Therefore, the 
Hamiltonian consists of three parts, 
\beq
H=T+V_i+V_{12}\,
\eeq
the kinetic energy of the electrons,
\beq
T=\frac{{\bf p}_1^2}{2}  + \frac{{\bf p}_2^2}{2} \,
\eeq
the potential energies associated with the interaction with 
the nucleus and the field (polarized along $z$-axis),
\beq
V_i = -\frac{2}{|{\bf r}_1|} -\frac{2}{|{\bf r}_2|}  
+ F(t) z_1 + F(t) z_2
\eeq
and the repulsion between the electrons,
\beq
V_{12} = \frac{1}{|{\bf r}_2 - {\bf r}_1|}\,.
\eeq
The electric field strength $F(t)$ has an oscillatory component
times the envelope from the pulse; the discussion applies
to general $F(t)$, and specific choices have to be made for the 
numerical simulations only. 

Once the electrons leave the atom the repulsion pushes them apart and becomes
weaker the larger the separation. Thus in the asymptotic state after ionization
and after the pulse is turned off, repulsion is minimized. In order to identify
the effects of the electron repulsion on the full process it is
instructive to consider the double ionization events without repulsion first.
This will be presented in the next section. The pathways with electron repulsion
included will be presented thereafter.

\subsection{Without electron repulsion}
The Hamiltonian for two electrons without electron repulsion splits
into two independent Hamiltonians for each electron. In view of the fast
motion of the electrons close to the nucleus we will frequently use
an adiabatic assumption and discuss motion of the electrons in
a field with fixed amplitude $F$. Note, however, that all simulations
use the full time-dependent field and do not make use of this
adiabatic assumption.
With this assumption each electron moves
in a constant electric field, one of the few
non-trivial integrable problems. The initial energy and the other constants
of motion of each electron are fixed by the initial conditions 
(in the compound state) and do not change. Double ionization
can thus occur only if both electrons individually have enough energy 
to ionize (and have the other constants of motion so as to allow
ionization). The threshold for ionization is set by the field strength:
if the field is non-zero a Stark saddle forms and the total energy
has to be above the Stark saddle. For a constant field strength
$F$ the saddle lies at $|z_F|=\sqrt{2/|F|}$ and has a potential energy
(single electron) $V_F=-2\sqrt{2|F|}$. Therefore, double ionization is 
excluded if the total energy for both electrons is less than
$2V_F$. For $E=2V_F$ the only path leading to double ionization has
both electrons with the same energy. For $E>2V_F$ double ionization 
becomes possible even with slight asymmetries in the energy distribution.

In the full system of two electrons without repulsion but in
a time-dependent field integrability is lost
but separability is still preserved. As discussed in the introduction
 the initial
state is a compound state
with negative energy for each electron. If that energy is 
above the Stark saddle and if the motion is directed towards it
the electrons can cross and run away from the nucleus. Once they cross
the saddle they have to gain energy to escape from the Coulomb attraction
when the field is turned off. This happens essentially by running down
the potential energy slope on the other side of the saddle while the field
is still on. This mechanism of energy gain is the same as in the 
interacting electron case.

\subsection{With electron repulsion}
With the electron repulsion included the common Stark saddle at
$z_F$ is no longer accessible since the electrons cannot sit on 
top of each other. The best that can be achieved is a symmetric
arrangement of both electrons in the same distance from the nucleus
and symmetric with respect to the field axis. If the distances are not
the same, electron repulsion will push the electron that is further out
away from the nucleus and thus help towards ionization, but the one further
in has to face not only Coulomb attraction but also the repulsion from
the one further out. Thus repulsion acts so as to amplify
differences in energy in this configuration. The configuration that is
singled out is a symmetric one, with both electrons moving 
at the same distance
from the nucleus.
Deviations from this configuration will be amplified 
sufficiently so that non-sequential double ionization is suppressed and only
single ionization takes place.
A remaining electron can, however, be still ionized if its energy is
higher than the saddle energy for a single electron atom and, in the adiabatic
picture, the other constants of motions allow to do so. In such 
sequential ionization correlations between escaping electrons are 
strongly weakened, 
that is the final momenta of the electrons along the polarization axis can 
be either parallel or anti-parallel.  

\section{Dynamics in the $C_{2v}$ and $C_{v}$ subspaces}

A two electron atom illuminated by a linearly polarized electromagnetic wave
possesses some 
symmetry subspaces. The simplest $C_{2v}$ symmetric configuration 
corresponds to
both electrons moving in a plane which contains the field polarization 
axis and with positions and momenta symmetric with respect to this 
axis. The electrons put in such a configuration never leave it
because there is no force which can pull them out of the subspace. 
The symmetry subspace can be enlarged. That is, with additional 
bending motion of the electrons with respect to the field axis the symmetry
subspace is $C_{v}$. 

The symmetric configurations become more apparent in suitably
chosen coordinates. We apply the canonical transformation
\bea
R=(\rho_1+\rho_2)/2,& p_R=p_{\rho_1}+p_{\rho_2} \cr
r=(\rho_1-\rho_2)/2,& p_r=p_{\rho_1}-p_{\rho_2} \cr
Z=(z_1+z_2)/2,& p_Z=p_{z_1}+p_{z_2} \cr
z=(z_1-z_2)/2,& p_z=p_{z_1}-p_{z_2}, \cr
\varphi=\varphi_1+\varphi_2,& L=(p_{\varphi_1}+p_{\varphi_2})/2 \cr
\phi=\varphi_1-\varphi_2,& p_\phi=(p_{\varphi_1}-p_{\varphi_2})/2 
\label{can}
\eea
where $(\rho_i,z_i,\varphi_i)$ are cylindrical coordinates of the electrons,
labeled $i=1$ and $2$.
For double ionization in linearly polarized laser field the total angular
momentum projection on the polarization axis is conserved.  
The experiments  begin with atoms in the ground state, thus, for the field
directed along the $z$-axis we may choose $L=0$. Then the Hamiltonian of 
the system reads
\bea
H&=&\frac{p_R^2+p_r^2+p_Z^2+p_z^2}{4}+\frac{p_\phi^2}{2(R+r)^2} \cr
&&+\frac{p_\phi^2}{2(R-r)^2}+V(R,r,Z,z,\phi,t)
\eea
with the potential energy 
\bea
V&=&-\frac{2}{\sqrt{(R+r)^2+(Z+z)^2}}-\frac{2}{\sqrt{(R-r)^2+(Z-z)^2}} \cr
&&+\frac{1}{\sqrt{2R^2-2(R^2-r^2)\cos\phi+2r^2+4z^2}} \cr
&&+2ZF(t),
\label{full}
\eea
where the field is given by
$F(t)=Ff(t)\cos(\omega t+\theta)$ with $F$, $\omega$ and $\theta$
the peak amplitude, frequency and initial phase of the 
field, respectively, and with
\beq
f(t)=\sin^2(\pi t/T_d)
\label{shape}
\eeq
the pulse envelope of duration $T_d$.

Setting $r=0$, $p_r=0$, $z=0$ and $p_z=0$ we define 
the $C_{v}$ symmetry subspace. The Hamiltonian is then reduced to
\beq
H=\frac{p_R^2+p_Z^2}{4}+\frac{p_\phi^2}{R^2}+V(R,Z,\phi)
\label{hcv}
\eeq
where
\bea
V&=&-\frac{4}{\sqrt{R^2+Z^2}}
+\frac{1}{R\sqrt{2(1-\cos\phi)}}+2ZF(t).
\label{pcv}
\eea
The potential (\ref{full}) is symmetric in $r$ (for $z=0)$ and in 
$z$ (for $r=0$),
so that the derivatives with respect to $r$ and $z$ vanish: once the
electrons are in the symmetry subspace $r=z=0$ and $p_r=p_z=0$, 
they cannot leave it.
\begin{figure}
\centering{\psfig{file=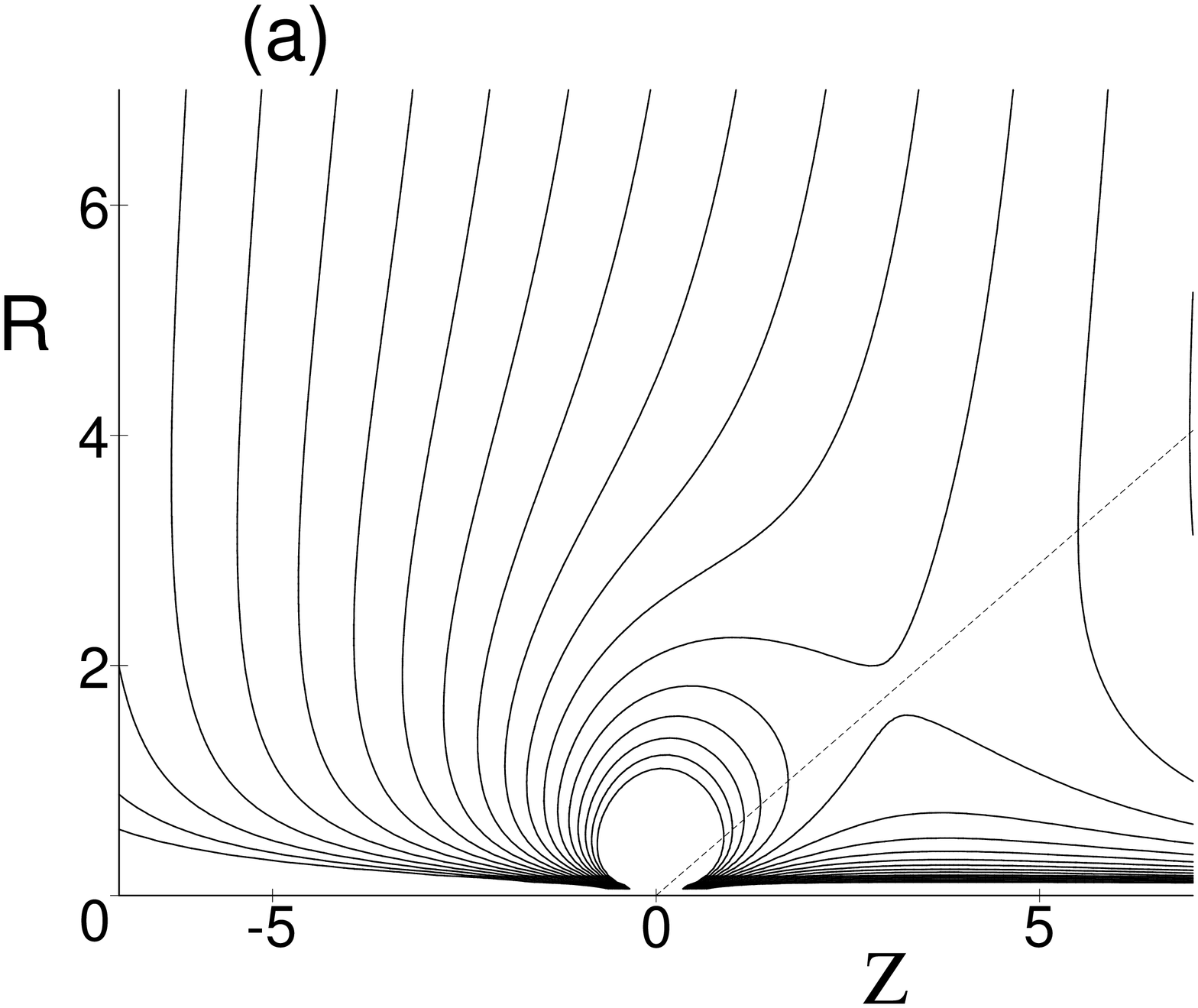,scale=0.3,angle=-90}
\psfig{file=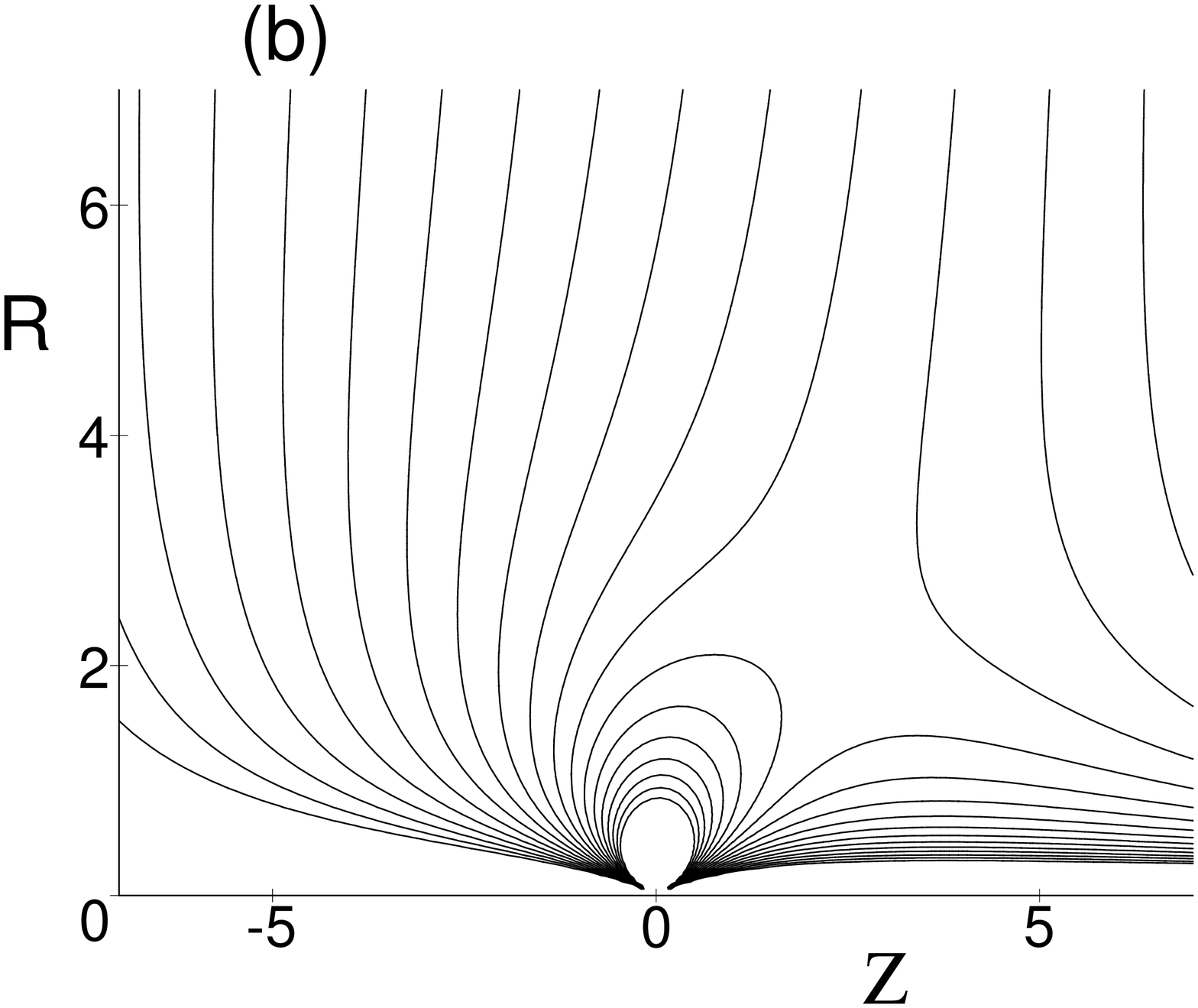,scale=0.3,angle=-90}}
\caption[]{
Sections through equipotential surfaces of the adiabatic 
potential Eq.~({\protect \ref{pcv}}) for fixed time $t$ and for $\phi=\pi$ (a)
and $\phi=\pi/4$ (b). Panel (a) corresponds also to the potential 
Eq.~({\protect \ref{pc2v}}) in the $C_{2v}$ symmetric subspace;
the saddle moves along the 
dashed line when the electric field points in the positive
$Z$-direction and along a second obtained by reflection on
$Z=0$ during the other half of the field cycle. 
}\label{potential}
\end{figure}

The further constraint $\phi=\pi$ and $p_\phi=0$
leads to the $C_{2v}$ symmetry subspace
\beq
H=\frac{p_R^2+p_Z^2}{4}+V(R,Z)
\eeq
with potential
\beq
V=-\frac{4}{\sqrt{R^2+Z^2}}
+\frac{1}{2R}+2ZF(t).
\label{pc2v}
\eeq

Let us begin with an analysis of the motion in the $C_{2v}$ subspace.
The electrons move in a plane and their positions ($\rho_i=R$, $z_i=Z$) 
in that plane and their momenta ($p_{\rho_i}=p_R/2$, $p_{z_i}=p_Z/2$) 
are the same.

The adiabatic potential (\ref{pc2v}) for fixed time corresponding to
the maximal field amplitude $F=0.137$~a.u., i.e. an intensity of 
$6.6\cdot 10^{14}$ W/cm$^2$, is shown in Fig.~\ref{potential}a. 
The saddle is located
along the line $Z_S=r_S \cos \theta_S$ and $R_S=r_S \sin \theta_S$ with
$\theta_S = \pi/6$ or $5\pi/6$ and at a distance
\beq
r_S^2=\sqrt{3}/|F(t)|\,.
\eeq
The energy of the saddle is 
\beq
V_S = - 6 \sqrt{|F(t)|/\sqrt{3}}\,.
\label{saddle}
\eeq
For the above mentioned field the saddle has an energy of 
$V_S=-1.69$~a.u..

If we switch off the repulsion between the electrons 
the saddle will move onto the $Z$-axis, i.e. both electrons are allowed 
to escape symmetrically on top of each other. 
With repulsion the saddle splits into two symmetrically related
ones and moves away from the $Z$-axis. 

\begin{figure}
\centering{\epsfig{file=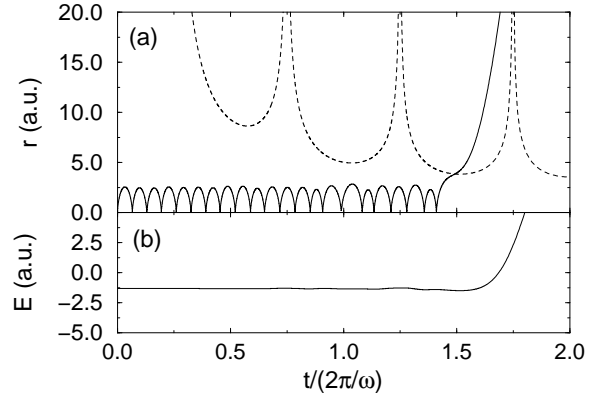,scale=0.35,angle=-90}}
\caption[]{
A typical trajectory in the $C_{2v}$ symmetric subspace with $E=-1.3\, $~a.u., 
$F=0.137$~a.u. and the pulse duration $T_d=4\times 2\pi/\omega$.
Panel (a) distance of the electrons to
the nucleus. The dashed line indicates the distance
of the saddle. Note that before the double ionization occurs the
effect of the field on the electrons is minimal, supporting the adiabatic
assumption.
Panel (b) energy of the electrons. 
Note that the initial state has negative total energy and cannot
lead to double ionization. The energy increases once the electrons
have escaped from the nucleus far enough so that acceleration by
the electric field dominates.
}\label{traj}
\end{figure}

A typical trajectory within the symmetric configuration 
for $\omega=0.057$~a.u. (800~nm)
is shown in
Fig.~\ref{traj}. During the ramping of the field the electronic motion
is little influenced by the electric field, but during the third half cycle
of the field the saddle is close enough to the electron orbits and 
ionization takes place. Once on the other side of the saddle,
the electrons rapidly gain energy.
The saddle thus provides a kind of transition state \cite{Wigner,Pollak}
for the correlated double ionization process: once the electrons cross it, 
they are accelerated by the field and pulled further away, making
a return rather unlikely. Moreover, the electrons can acquire the missing 
energy so that both can escape to infinity even when the field vanishes. 
Note that before double ionization occurs the
effect of the field on the electrons is minimal, supporting the adiabatic
assumption.

In the experiments \cite{weber1,weber2,rottke,weber3}
ion momenta both parallel and transverse to the field are
measured. For $\omega=0.057$~a.u. momentum transfer by photons is negligible,
so the ion momentum reflects the sum of the momenta of the emitted electrons,
${\mathbf p}_1+{\mathbf p}_2=-{\mathbf p}_{ion}$. Symmetric
motion in the $C_{2v}$ subspace
takes place in a plane and consequently the total transverse momentum 
of the electrons is zero. In the $C_v$ subspace the electrons are permitted to
leave the plane, and this is a minimal extension necessary to
give obtain non-vanishing transverse momenta. 

The sections through equipotential surfaces 
of Eq.~(\ref{pcv}) for $\phi=\pi$ and $\phi=\pi/4$ 
are shown in Fig.~\ref{potential}. Increasing or decreasing 
$\phi$ from $\pi$ results in greater repulsion energy between the electrons. 
So the electrons can escape, with
the coordinate $\phi\ne\pi$, only if the energy is greater than the saddle
energy, Eq.~(\ref{saddle}). 

We discuss here double ionization of atoms but the analysis of a 
symmetric escape can be easily extended to a multiple ionization process.
Especially for $N$ electrons the symmetric subspace corresponding to $N$
particles symmetrically distributed in the plane perpendicular to the
polarization axis is $C_{Nv}$ \cite{ES2}. 

\section{The saddle}
Within the adiabatic assumption and for a fixed external
field strength non-sequential double ionization hinges
on the crossing of the saddle in the $C_{2v}$ subspace. For the dynamics
within the subspace this is obvious from the potential. For the motion
in the full six-dimensional configuration space this is less clear. One
possibility to gain insight locally near the saddle is a harmonic
analysis and a determination of the frequencies of small deviations.
Within the $C_{2v}$ symmetric subspace there is one hyperbolic mode 
corresponding
to motion across the saddle (the `reaction coordinate', in
chemical physics parlance \cite{Wigner,Pollak}) 
and a stable one corresponding to perpendicular variations.
In the full space we expect at least one additional unstable one, 
corresponding to the amplification of energy differences mentioned before
in section (IIB).
The analysis in this section is for fixed field strength, for electron
dynamics in a constant external field, justified by the 
adiabatic reasoning. 

The second order variations of the potential (\ref{full}) around the saddle 
point results in
\bea
H&\approx &\frac{p_R^2+p_Z^2}{4}+\frac{9}{2r_S^3}(R-R_S)^2 \cr
&&-\frac{3\sqrt{3}}{r_S^3}(R-R_S)(Z-Z_S)-\frac{5}{2r_S^3}(Z-Z_S)^2 \cr
&& +\frac{p_r^2+p_z^2}{4} 
+\frac{1}{2r_S^3}r^2-\frac{3\sqrt{3}}{r_S^3}rz-\frac{9}{2r_S^3}z^2 \cr
&& +\frac{p_\phi^2}{R_S^2}
+\frac{1}{4R_S}(\phi-\pi)^2+V_S.
\eea
The $\phi$-degree of freedom corresponds to a bending motion of 
the electrons against each other and is stable on account of the 
repulsive nature of the Coulomb force. This degree of freedom is
also the one that comes in by going from the $C_{2v}$ subspace to 
the $C_v$ subspace.

Diagonalization in the ($R-R_S,Z-Z_S$) space reveals one
stable and one unstable mode. The latter corresponds to the reaction
coordinate and its Lyapunov exponent is $\mu\approx 1.21F^{3/4}$. 
Similar analysis in 
the ($r,z$) space yields another stable and unstable mode with Lyapunov
exponent $\nu\approx 1.57F^{3/4}$. The direction of the unstable mode is 
$(w_r,w_z)\approx (0.43w,w)$ and it
corresponds to the situation when one electron escapes and the other one is
turned back to the nucleus. That is, with positive and increasing $w$ 
the first electron moves away from the nucleus, i.e. $\rho_1=R_S+w_r$ and
$z_1=Z_S+w_z$ grow [see (\ref{can})], while the other one returns
to the nucleus, i.e. $\rho_1=R_S-w_r$ and $z_1=Z_S-w_z$ decrease.
 
All in all there are three stable modes and two unstable ones. 
Any energy contained in the stable modes is preserved and cannot
be transferred to kinetic energy along the reaction coordinates.
Whether single or double ionization occurs is thus determined solely by
the energy and initial conditions in the two hyperbolic subspaces.
For energy equal to the saddle energy only a trajectory within the 
$C_{2v}$ symmetry
subspace leads to non-sequential escape -- any deviation from the subspace are
growing faster than the escape along the reaction coordinate since 
$\nu>\mu$. For energy higher than the saddle some deviations from the symmetry 
subspace are allowed.
In particular, following Wanniers lead 
and Rosts generalization \cite{Wannier,Rau,Rost,Rost1} it
is possible to estimate the critical behavior for the double ionization
cross section at threshold. It is algebraic with the exponent given by
the ratio of the positive Lyapunov exponents of the unstable modes.
A detailed discussion of this is outside the main line of 
our arguments here and will be given elsewhere.

It is instructive to actually calculate numbers for the 
Lyapunov exponents in the two directions. For lasers with the 
maximal field strength of $F=0.137$~a.u.
we find $1/\mu=3.7$~a.u. and $1/\nu=2.8$~a.u.. 
Compared to the period of the laser, $2\pi/\omega=110.2$~a.u. 
this is rather fast, indicating that the crossing of 
the saddle and the separation away from the double ionization
manifold take place rather quickly. This justifies also 
our adiabatic analysis in this section.

\section{Final state momenta distribution within the symmetric subspaces}

So far we have discussed the phase space features in an adiabatic
approximation for fixed field strengths. Now we will use this to 
draw conclusions about the experimentally observed signatures, specifically
about the distributions of ion momenta in the final state. 
They can be calculated within the $C_v$ subspace 
by averaging over all initial conditions and all phases of the field.
That this is possible is connected with the instability of the saddle:
all trajectories leading to the non-sequential ionization 
have to pass sufficiently close to the saddle and the symmetric subspace.
It therefore is possible to estimate the behaviour near the subspace
from the dynamics within the subspace.

\subsection{Parameters of the model}

The rescattering of an electron leads to a highly excited complex
of total energy $E$ which is one of the parameters of our model.
The maximal energy a rescattering electron can bring in has
been estimated to be about 
$3.17U_p$ \cite{Corkum,Kulander3}, where $U_p$ is the ponderomotive 
energy of an electron. 
For the weakest field used in the experiment on 
double ionization of He atoms \cite{weber1} this maximal energy barely 
corresponds to the ionization energy of the He$^{1+}$ ions. 
We therefore assume in the following
that the highly excited complex has a negative 
initial energy, $E<0$. 

The absence of detailed knowledge of the structure of the initial
compound nucleus suggests to average over the initial configurations.
However, even for a negative 
energy and fixed time it is difficult to define a microcanonical
distribution of initial conditions for the Hamiltonian (\ref{hcv})
since, for non-zero external field, the
system is open. Therefore, we choose for the calculations
initial conditions
from the energy shell that also lie in the hypersurface $Z=0$. 
The results are not
sensitive to a particular choice of the hypersurface but the one for $Z=0$
has the advantage that the dipole moment along the filed is zero and the choice
of the initial conditions does not depend on the 
initial field phase.

\begin{figure}
\centering{\epsfig{file=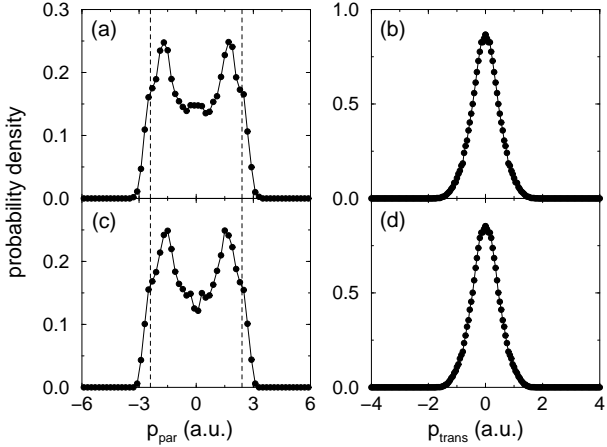,scale=0.33,angle=-90}}
\caption[]{
Final distributions of ion momenta parallel, $p_{par}$ and transverse,
$p_{trans}$ to the field polarization axis 
for the initial energy $E=-0.1$~a.u., peak field amplitude 
$F=0.137$~a.u. and pulse duration $T_d=20\times 2\pi/\omega$.  
Panels (a)-(b) correspond to the initial time $t_0=0.25T_d$ in the pulse
duration with the envelop $f(t)=\sin^2(\pi t/T_d)$ while
panels (c)-(d) to $t_0=0.75T_d$. Dashed lines are related to the estimates
$\pm2Ff(t_0)/\omega=\pm2.4$~a.u.. Note that the
distributions are essentially the same independently if one chooses 
$t_0$ before
or after the peak field value provided $f(t_0)$ is the same.
The results are based on integrations of
about $8\cdot10^4$ 
trajectories. 
}
\label{dwidth1}
\end{figure}

\begin{figure}
\centering{\epsfig{file=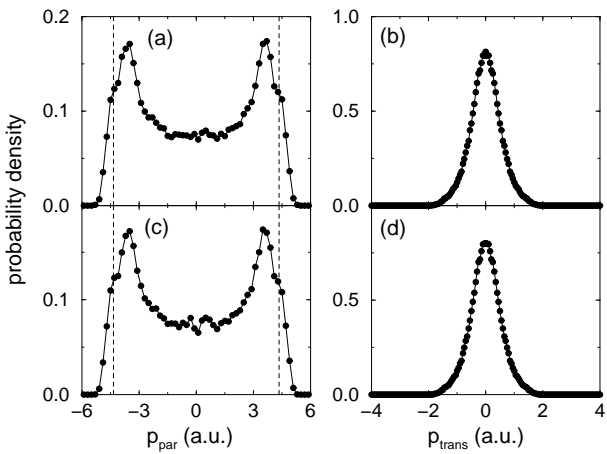,scale=0.33,angle=-90}}
\caption[]{
The same as in Fig.~{\protect \ref{dwidth1}} but for $t_0=0.4T_d$ [panels
(a)-(b)] and $t_0=0.6T_d$ [panels (c)-(d)]. The width of the parallel momentum
distribution can be estimated as $\pm2Ff(t_0)/\omega=\pm4.3$~a.u..
}
\label{dwidth2}
\end{figure}

The second parameter, in addition to the energy, is the time 
$t_0$ during the pulse
Eq.~(\ref{shape}) when the highly excited complex is formed. The rescattering
event is not possible at the beginning of the pulse, so one has to start
simulations somewhere in the middle of the pulse.
In Fig.~\ref{dwidth1} and \ref{dwidth2} final distributions of ion momenta 
for the
initial energy $E=-0.1$~a.u., field strength $F=0.137$~a.u. and different
initial time $t_0$ are shown. The distributions of the transverse momenta 
are almost the same but the ones for the parallel momenta differ. The latter
reveal a double hump structure with widths sensitive to the
initial time. 

The maximum energy that can be acquired by a free electron 
in the field is $2U_p$. So, for
parallel emission of two electrons, the maximal parallel 
ion momentum can be estimated as
\beq
p_{par}=2\sqrt{4U_p}=2Ff(t)/\omega.
\label{ionmax} 
\eeq
If we substitute in Eq.~(\ref{ionmax}) $t=t_0$ we find values which 
correspond very well to the widths of the distributions in 
Fig.~\ref{dwidth1} and \ref{dwidth2}. The figures 
indicate also that the widths are the same
independently if $t_0$ is chosen before or after the peak of the pulse provided
$f(t_0)$ is the same. 
This implies that the dominant ionizations take place in
the first field cycle after the complex is formed.
Fitting the width of the calculated distribution to the experimental results
allows one to estimate the moment in the pulse when majority of doubly ionized 
ions are created.
\begin{figure}
\centering{\epsfig{file=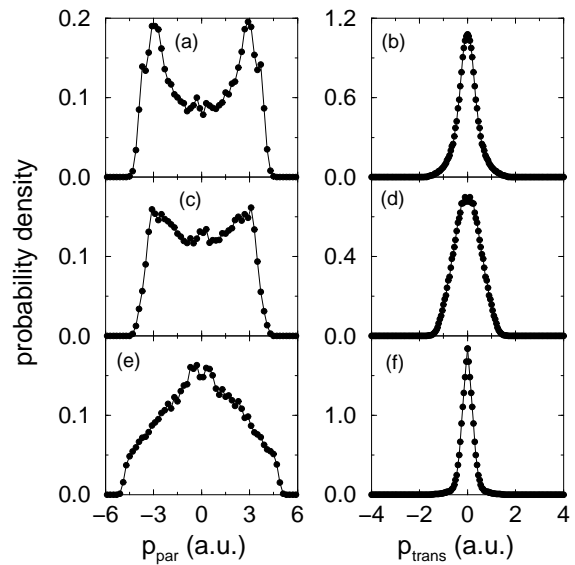,scale=0.4,angle=-90}}
\caption[]{
The same as in Fig.~{\protect \ref{dwidth1}-\ref{dwidth2}} 
but for fixed $t_0=0.33T_d$ and different initial energy: $E=-0.05$~a.u. 
[panels (a)-(b)], $E=-0.5$~a.u. [panels (c)-(d)] and 
$E=-1.5$~a.u. [panels (e)-(f)].
}
\label{dshape}
\end{figure}

Now we fix $t_0$ and change the initial energy $E$; the results are
shown in Fig.~\ref{dshape}. For the lowest energy, $E=-1.5$~a.u., 
the transverse momentum distribution is narrower than for higher energies.
$E=-1.5$~a.u. is actually very close to the 
minimal saddle energy for the applied field, $V_S=-1.69$~a.u.. Thus
the effect is natural as, close to the saddle energy, only trajectories near the
$C_{2v}$ subspace can cross the saddle and those with $\phi\ne\pi$ bounce back
from the potential barrier and do not ionize. 

While the width of the distributions of the parallel momenta do not change
significantly their shapes do, especially for the initial energy close to the
saddle. The electrons that cross the saddle when the energy is near $V_S$ 
are slow and the combined interactions of the external and
Coulomb fields shapes the distributions. This is different from the high energy
case: then 
shortly after the electrons cross the saddle the interaction with the 
electric field is stronger than the attraction to the nucleus and the
distribution is mostly shaped by the laser field. 

The initial energy of the complex is the higher the higher intensity of the
laser is and the larger the energy a rescattered electron can bring in
\cite{Corkum,Kulander3}.
{}From the dependence of the distributions on the initial energy 
we may conclude 
that in the experiment the shape of the distribution of the parallel ion momenta
should change character when the laser intensity increases. For the
intensity at the threshold for non-sequential double ionization the distribution
with single maximum around zero momentum is expected; for higher intensities the
double hump structure should turn up.

All numerical results have been 
obtained for initial conditions taken from the $C_v$
symmetry subspace. Our results for the $C_{2v}$ subspace for parallel momentum
distributions are essentially the same.
The transverse momenta of ions for the $C_{2v}$ subspace are, however, 
zero because of the symmetry assumption.

After this discussion of the two parameters (initial energy and 
starting time of the integration) we can turn to comparisons with experimental
observations.

\subsection{Comparison with experimental results}

\begin{figure}
\centering{\epsfig{file=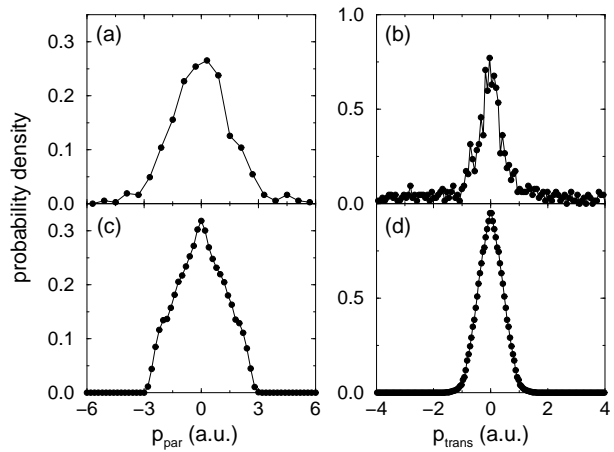,scale=0.33,angle=-90}}
\caption[]{
Panels (a)-(b): final ion momentum distributions measured in the experiment of
double ionization of He atoms in the focus of 800~nm, 220~fs (i.e. about
$80\times 2\pi/\omega$) laser pulses at peak intensity of 
$2.9\times 10^{14}$~W/cm$^2$ (i.e. for the field strength $F=0.091$~a.u.) from
{\protect \cite{weber1}}.
Panels (c)-(d): the corresponding distributions calculated in the $C_v$
symmetry subspace for the initial energy $E=-0.6$~a.u. and $t_0=0.33T_d$ where
$T_d/2=80\times 2\pi/\omega$, see Eq.~{\protect \ref{shape}}.
}
\label{he1}
\end{figure}

\begin{figure}
\centering{\epsfig{file=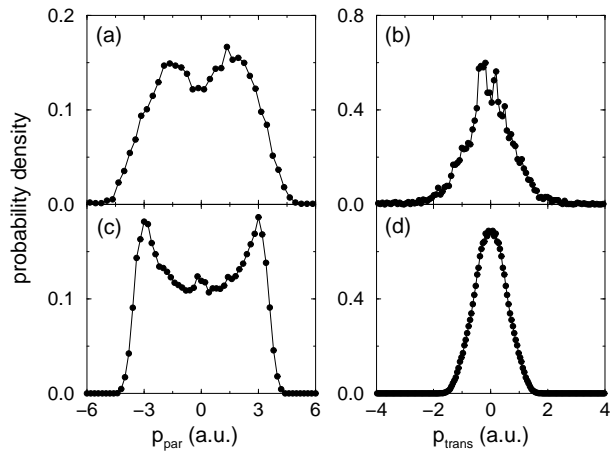,scale=0.33,angle=-90}}
\caption[]{
Panels (a)-(b): the same as in the corresponding panels in Fig.~{\protect
\ref{he1}} but for the peak intensity of 
$6.6\times 10^{14}$~W/cm$^2$ (i.e. for the field strength $F=0.137$~a.u.).
Panels (c)-(d): the same as in the corresponding panels in Fig.~{\protect
\ref{he1}} but for the initial energy $E=-0.4$~a.u..
}
\label{he2}
\end{figure}

Weber {\it et al.} \cite{weber1}
carried out double ionization experiments with He
atoms and measured the distributions of ion momenta. They applied 
infrared (800~nm) laser pulses of the duration 220~fs 
(measured on the half peak
value) and with the peak intensities in the range
$(2.9-6.6)\times 10^{14}$~W/cm$^2$. 
In Fig.~\ref{he1} and \ref{he2} we show the
experimental distributions and compare with those calculated in the $C_v$
subspace. The agreement is very good except for the parallel momentum
distribution in Fig.~\ref{he2} where the calculated distribution possesses 
a much more pronounced minimum than in the experiment and the positions of 
the peaks do not exactly coincide with the experimental values. 

There are a few possible sources for these discrepancies. 
First, the pulse duration in the experiment was 
quite long, i.e. about 80 field cycles. The slow ramping 
of the field in the experiment implies that the initial time $t_0$ of
the ionization is less well defined, i.e. there are contributions from some
range of $t_0$. There are also contributions from different initial energies.
Secondly, real ionizing trajectories do not live exactly 
in the symmetry subspace but
close to it, leading to asymmetries and 
additional differences in the final momenta.
And there are also possible contributions from sequential double
ionization events (see below).

\begin{figure}[hbt]
\centering{\epsfig{file=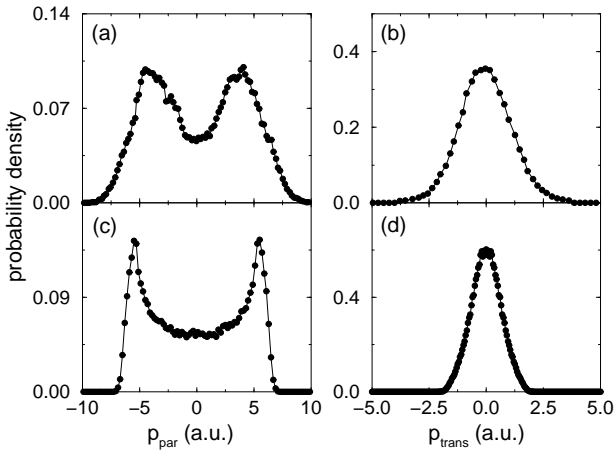,scale=0.33,angle=-90}}
\caption[]{
Panels (a)-(b): final ion momentum distributions measured in the experiment of
double ionization of Ne atoms in the focus of 795~nm, 30~fs (i.e. about
$11\times 2\pi/\omega$) laser pulses at peak intensity of 
$13\times 10^{14}$~W/cm$^2$ (i.e. for the field strength $F=0.192$~a.u.) from
{\protect \cite{rottke}}.
Panels (c)-(d): the corresponding distributions calculated in the $C_v$
symmetry subspace for the initial energy $E=-0.3$~a.u. and $t_0=0.4T_d$ where
$T_d/2=11\times 2\pi/\omega$, see Eq.~{\protect \ref{shape}}.
}
\label{ne}
\end{figure}

Moshammer {\it et al.} \cite{rottke}
performed experiments with Ne atoms for much shorter
pulses, i.e. 30~fs and for radiation with similar wave length (795~nm) as the
previous group. The comparison of our calculations with this experiment are
presented in Fig.~\ref{ne}. Interactions of the 
two excited electrons with the other electrons are neglected in our model and 
the energy values used in the calculations are measured with respect to the 
threshold for the two electron continuum (i.e. about 2.3~a.u.). 
The agreement is even better than for the case of He atoms. 
This is presumably due to the much shorter pulse duration 
and the fasted ramping of the field, so 
that the time $t_0$ when the majority of the ionization events happen is 
much better defined.

\section{Sequential double ionization}

Already from the experiment it is clear that double ionization
is a rare process, e.g. outnumbered by single ionization by about
$10^4:1$ for He atoms and field intensity $2.9\cdot 10^{14}$~W/cm$^2$
\cite{weber1}. 
Hence arbitrarily chosen initial conditions in
the full space will typically not lead to double ionization and numerical
simulations of the whole process are rather unattainable.

\begin{figure}
\centering{\epsfig{file=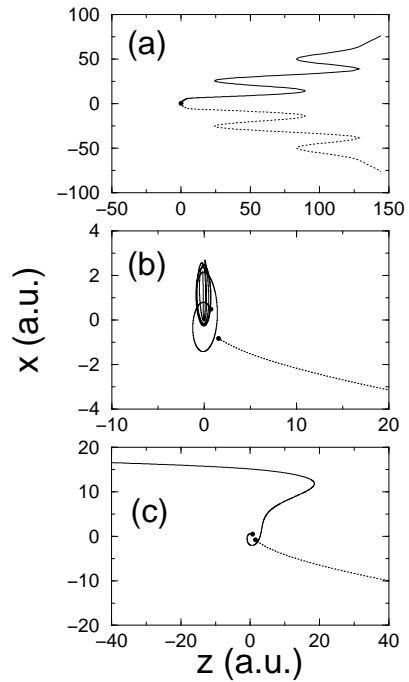,scale=0.5,angle=-90}}
\caption[]{
Trajectories of electrons outside the symmetric subspace
for $E=-0.58$~a.u. and $F=0.137$~a.u.. Initial 
positions are close to the saddle and marked by heavy dots; the electrons
are distinguished by dotted and continuous tracks. 
Panel (a) shows a symmetric escape of the electrons.
Panel (b) shows a case where outside the symmetry subspace one
electron escapes and the other falls back to the ion. Panel (c)
shows an example of sequential ionization of both electrons in
opposite directions.
}
\label{3dtraj}
\end{figure}

We have discussed the non-sequential double escape of the electrons considering
trajectories within the symmetry subspace. Motion in the symmetry subspace is
unstable, that is deviations from the subspace will be amplified leading to
single rather than double ionization. We can
illustrate this with trajectories started slightly
outside the symmetry plane (Fig.~\ref{3dtraj}). Fig~\ref{3dtraj}a shows 
initial conditions on the saddle and symmetrically escaping
electrons. For some deviation from symmetry, one electron escapes,
the other remains trapped to the nucleus (Fig.~\ref{3dtraj}b). 

It is possible, however, that the second electron returns to
the nucleus but picks up enough energy 
to ionize itself (Fig.~\ref{3dtraj}c). In the adiabatic picture, 
if the energy of the remaining electron is higher than the saddle for a single 
electron atom, $V_F=-2\sqrt{2|F|}$, whether the electron stays trapped
or escapes depends on the other constants of motion (besides the
energy) in this integrable system.
The second electron can thus escape during the same half
field cycle as the first one but 
its final parallel momentum component need not be related to
that of the first electron, as is shown in Fig.~\ref{3dtraj}c. 

\begin{figure}
\centering{\epsfig{file=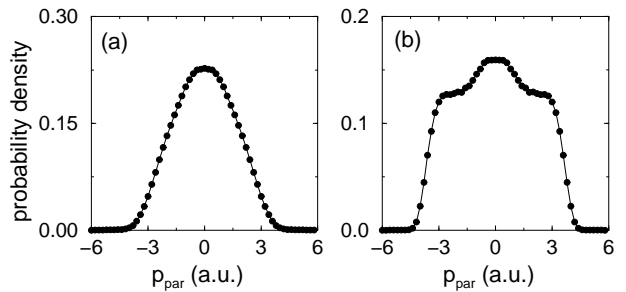,scale=0.35,angle=-90}}
\caption[]{
Final parallel ion momenta distributions calculated in the two-dimensional 
non-interacting electrons model for the initial time $t_0=0.33T_d$ in 
the pulse duration, where $T_d=20\times 2\pi/\omega$, peak field amplitude 
$F=0.137$~a.u. and 
initial energy $E=-0.8$~a.u. [panel (a)] and $E=-0.1$~a.u. [panel (b)].
The initial conditions for the electrons have been chosen to satisfy
$E=E_1+E_2$ with the restrictions $E_1<0$ and $E_2<0$.
The saddle energy for a single electron is $V_F=-1.05$~a.u..
The results in each panel are based on about $1.5\cdot 10^6$ trajectories.
}
\label{nonint}
\end{figure}
 
We are not able to calculate the contribution of sequential events 
to the double ionization process but we can estimate 
the distribution of the final ion momenta if the sequential process 
were dominant. 
To simulate sequential escape one may return to a non-interacting electron
model, essentially since the electrons cross the barrier at different times. 
In the model the initial conditions for each electron are chosen
independently, constrained only by the requirement that the total initial 
energy is fixed $E=E_1+E_2$ and that $E_1<0$ and $E_2<0$. 
Simultaneous double ionization are not explicitly excluded, but events with
delayed ionization are more probable, so that the distributions can
still reflect the contributions from sequential ionization.

In Fig.~\ref{nonint} the distributions of parallel ion momenta for
$E=-0.8$~a.u. and $E=-0.1$~a.u., calculated in the non-interacting electrons
model for $F=0.137$~a.u, are plotted. The figure should be
compared with the figures from the previous section.
The conclusion is straightforward: the non-interacting electrons model is 
not able
to reproduce correlations between the electrons observed in the experiments.
Moreover, sequential escape can not be a dominant mechanism for double 
ionization in the range of the field intensities considered here.

\section{Conclusions}

We have considered the process of double ionization of atoms in a strong, 
linearly polarized field for intensities below the saturation of single
electron ionization.
We have developed a model for non-sequential
double ionization within classical mechanics. The process has been divided into
two stages: in the first one a rescattering process leads to a highly 
excited complex of two electrons. 
In the second stage, an ionization of such 
compound state takes place. 
We have focused on the  latter stage considering
different pathways to double escape of the electrons. 

The excited complex can doubly ionize even when its energy is negative because
the external field opens up saddles for electron escape. 
The pathway favored by the Coulomb interactions and the field
is simultaneous symmetric escape of both electrons. Deviations form the
symmetric configurations are amplified by the repulsion between the electrons
which pulls one electron to infinity but the other one is pushed back to the
nucleus. Therefore we propose that non-sequential double ionization is
dominated by motions of electrons in or near the symmetric
subspace with the saddle. 
 
The requirement of the symmetric motion 
greatly simplifies the analysis which then can be carried out for the 
three- or even two-dimensional effective potential.   
The trajectory simulations within the symmetric configurations turns out to
reproduce the experimentally observed ion momenta distributions very well.  
We have also considered an alternative mechanism of the ionization, i.e. 
sequential escape of the electrons. By means of the non-interacting electrons
model we show that the sequential ionization is not able to explain the
experimentally observed electrons correlations. 

The modeling of the experimental distributions requires information on
two parameters, the initial energy and the time of formation, 
which reflect a lack of knowledge on the 
compound state and the ramping of the field.
The dependence of the momentum distributions on the
parameters and comparison with the experimental results 
give insights into the dynamics of double ionization. 

The analysis in the present paper 
has been restricted to  double ionization but its extension to
multiple escape is straightforward \cite{ES2}. In its current form
the model is applicable for linearly polarized fields only. 
For other polarizations the number of rescattering events 
is greatly reduced. However, for some
elliptically polarized field, if an electron is driven back to the core and 
a highly excited complex is formed, 
in the adiabatic approximation the symmetric
configuration of the electrons can be defined with respect to the temporary 
electric field axis.   
Then one can proceed with the analysis as for the
linearly polarized case.

Our whole discussion has a more than superficial similarity
with Wanniers analysis of double ionization by electron impact
\cite{Wannier,Rau,Rost}.
The main differences are the presence of a field and its
time dependence, which enlarges phase space and complicates
the identification of the transition state. In the adiabatic
approximation at fixed field strength we could identify this
saddle in the $C_{2v}$ subspace. The comparison with
experiments is complicated furthermore by the necessity to
average over initial energy and time of preparation of the compound
state. Thus, signatures one might attribute to Wanniers
analysis, such as threshold exponents (they follow immediately
from the stability analysis of the saddle, but are not 
easy to verify), will be
even more difficult to study. But we have no
doubt that the observation of the correlated escape of the electrons
\cite{weber3} is a clear signature of the existence and dominance
of the saddle and the pathways to double ionization which we discuss
here.

\section{Acknowledgements}
We would like to thank Harald Giessen for stimulating our interest
in this problem and for discussions of the experiments.
Financial support by the Alexander von Humboldt
Foundation and by KBN under project 2P302B00915
are gratefully acknowledged.



\end{multicols}
\end{document}